\documentclass[twocolumn,10pt,a4paper]{article}
\usepackage{palatino}
\usepackage{amsmath}
\usepackage{amssymb}
\usepackage{graphicx}
\usepackage{graphics}

\newcommand{\Name}[1]{{#1}, }
\newcommand{\Review}[1]{#1, }
\newcommand{\Vol}[1]{#1}
\newcommand{\Year}[1]{#1 }
\newcommand{\Page}[1]{#1 }
\newcommand{\Book}[1]{#1}
\newcommand{\Publ}[1]{(#1)}

\newcommand{\REVIEW}[4]{\Review{#1} \Vol{#2} \Year{#3} \Page{#4}}

\title{Identifying Complex Networks by Random Walks}
\author{Filipi Nascimento Silva and Luciano da Fontoura Costa\inst{1}}

\author{
    Filipi Nascimento Silva and Luciano da Fontoura Costa \footnote{Cybernetic Vision Research
    Group, GII-IFSC, Universidade de S\~ ao Paulo, S\~{a}o Carlos, SP,
    Caixa Postal 369, 13560-970, Brasil, luciano@if.sc.usp.br.  }
    }

\begin{document}

\maketitle
\abstract{The possibility to identify the nature (e.g. random or
scale free) of complex networks while performing respective random
walks is investigated with respect to autonomous agents based on
Bayesian decision theory and humans navigating through a
graphic-interactive interface. The results indicate that the type of
the network (choice between random and scale free models) can be
correctly estimated in most cases.}

\vspace{1cm}
\emph{`They rebuild Ersilia elsewhere.  They weave a similar
pattern of strings which they would like to be more complex and at the
same time more regular than the other. (I. Calvino, Invisible Cities'}

\section{Introduction}

Conscious human existence can be understood as a trajectory in the
space-time phase space, unfolding as a consequence of our perceptions,
decisions and actions.  As suggested
recently~\cite{Luciano_know_walk,Luciano_rand_walk}, the dynamic
evolution of the life of a human individual can be approximated by a
random walk~\footnote{The term
\emph{random} is here meant to express general probabilistic
decision models, not necessarily the traditional random walk where
decisions are taken by uniformly sampling among the paths emanating
from each
node.}~\cite{Luciano_know_walk,Luciano_rand_walk,rand_walk1,
rand_walk2, rand_walk3,rand_walk4} in a complex network $\Gamma$
(e.g.~\cite{newman_survey,barabasi_survey,luciano_survey}) where the
nodes correspond to the possible decisions and the links to the
transitions between such decisions.  Note that, in case time is taken
explicitly, such a complex network will correspond to a decision tree
and include no cycles (i.e. a closed path).  In order to allow the
formation of cycles, we henceforth consider the evolution of time
implicitly, allowing that oneself will find her/himself in recurring
situations (e.g. choosing shoes for dinner).  A series of interesting
insights and results about our perception of the complexity of our
individual life (as far as decisions are concerned) can be achieved by
considering such a model.  For instance, in case the complex network
$\Gamma$ is scale free, the average degree of the nodes as sampled by
a traditional random walk will tend to result twice as large as its
real value (e.g.~\cite{Luciano_rand_walk}).  This interesting
phenomenon is a direct consequence of the fact that hubs are more
likely to be visited during a random walk in a BA network, hence the
overestimation of the average degree.

An important issue related to modeling human experience in terms of
random walks in complex networks concerns our ability to identify the
most likely mathematical model (e.g. Erd\~os-R\'enyi or
Barab\'asi-Albert) of the complex network being explored while
navigating on it through random walks.  The current article explores
this key issue from the perspective of having human subjects to
navigate through ER and BA complex networks models while trying to
identify their type.  In addition, in order to gain deeper insight on
this problem, the subjects correct ratio is compared with results
obtained by a agent that uses a optimal identification algorithm
(Bayesian decision).

The article starts by presenting basic concepts in complex networks
and follows by summarizing some statistical methods (Pearson
correlation coefficient and Bayesian decision theory) and explaining
the basic concepts of agent classification of networks. The article
follows by describing the experimental methodology and presenting its
results and comparison with the autonomous agents results.

\section{Networks Generation and Degree Distributions}

\begin{figure}
 \begin{center}
  \includegraphics[scale=0.3]{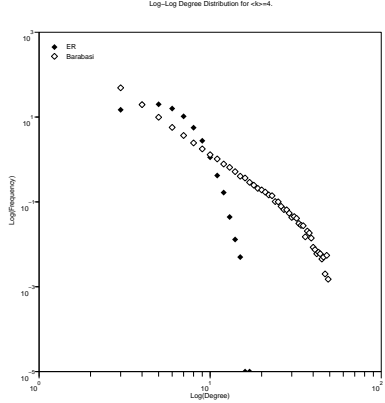}
  \caption{Average degree distributions for 2000 realizations of 
  ER and BA networks with average degree $\left< k \right>=4$. 
  ~\label{fig:distrib}}
 \end{center}
\end{figure}

Consider a matrix $K$, with $K(i,j)=K(j,i)=\{1~or~0\}$(i.e. a binary
and symmetric matrix). A non-weighted and non-oriented network (i.e. a
graph) can be completely specified by such an \emph{adjacency matrix},
where the existence of a connection between two nodes ($i$ and $j$) is
represented as $K(i,j)=K(j,i)=1$.  The adjacency matrix of a random
network (i.e. ER model) of size $N\times N$ can be generated by
starting with all elements equal to zero and making $K(i,j)=K(j,i)=1$
with probability $\rho$ for every pair of nodes $i$ and $j$, implying
average degree equal to $\left< k \right>=\rho N$.  The average degree
distribution obtained for 2000 realizations of the ER model with
$N=100$ and $\left< k
\right> =4$ is shown in Figure~\ref{fig:distrib}.

BA networks can be generated by selecting $M_0$ rows or columns,
$K(i,...)$ (initial nodes), and then connecting $e_0$ of these nodes,
randomly chosen, with a new node by filling the corresponding value on
adjacent matrix~\cite{newman_survey}. This process is repeated $t$
times, always connecting $e_0$ of previous selected nodes with a new
node.

Therefore, the final number of nodes will be $N=t+M_0$ and the number
of edges $e=e_0 t$, with average node degree given as $\left< k
\right>=\frac{2 e_0 t}{(M_0+t)}$.  For values of $M_0<<t$, the average
degree will be $\left< k \right>=2e_0$. Note that for low values of
$M_0$ the BA models can only be generated with even average degree.

The degree distribution of BA models is known to follow a
power-law~\cite{newman_survey,barabasi_survey,luciano_survey} as a
consequence of its scale-free nature, with power coefficient $\alpha =
-3$. The average degree distribution obtained for 2000 realizations of
the BA model with $N=100$ and $\left< k
\right> =4$ is shown in Figure~\ref{fig:distrib}.

\section{Statistical Concepts}

Two basic statistical methodologies are used in the present work in
order to construct an agent that classify a network. Because of the
linear behavior of the distribution of average degree for BA model and
non-linear for ER model, this property can be exploited as a parameter
for the segregation between the two models, this can be made by using
the Pearson correlation coefficient as follows.

The Pearson correlation coefficient is a statistical measurement
quantifying how strong is the linear joint variation between two
random variables~\cite{book:correlation}. Given the normalized
distribution of two random discrete variables, X and Y, the Pearson
correlation coefficient between them is defined by the covariance of
that two variables divided by their standard deviation, i.e.:

\begin{equation}
\label{eq:pearson}
  r_{X Y}=\frac{cov(X,Y)}{\sigma_X \sigma_Y}
\end{equation}

Given $n$ samples of the random variables $X$ and $Y$, henceforth
expressed $x_i$ and $y_i$, the respective Pearson Correlation
Coefficient can be estimated as:

\begin{equation}
\label{eq:pearson2}
  r_{X Y}=\frac{\sum (x_i - \bar{x})(y_i - \bar{y})}{(n-1)\sigma_X \sigma_Y}
\end{equation}

Where $\bar{x}$ and $\bar{y}$ are the average values of elements $x_i$
and $y_i$, and $\sigma{x}$ and $\sigma{y}$ are the respective standard
deviations. These conditions bond the Pearson correlation coefficient
between -1 and 1. Values of $r_{X Y}$ near $0$ suggest absence of
linear correlation between the two variables, while values around $1$
and $-1$ indicate correlated or anti-correlated behavior,
respectively.  Note that a nearly straight distribution of points is
observed for the cases characterized by absolute values of Pearson
coefficients nearly equal to 1.

\begin{figure}
 \begin{center}
  \includegraphics[scale=0.3]{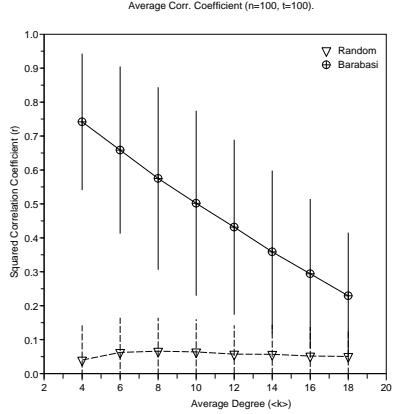}
  \caption{Average correlation coefficients for BA and ER models,
  taken from 2000 networks and considering every node of these
  networks.~\label{fig:meancorr}}
 \end{center}
\end{figure}

Because of the linear behavior of the logarithmic node degree
distribution observed for scale-free networks, contrasting with the
binomial distribution for ER, their squared correlation coefficients
can be used as a sound criterion for discrimination and identification
of these two models.  Therefore, the Pearson correlation coefficient
is used in this work in order to quantify the degree of straightness
of the log-log degree distribution. As shown in figure
~\ref{fig:meancorr}, the squared correlation coefficients for the
logarithmic of node degree distributions for BA model tend to be
substantially higher than those for the ER model. Note that the
difference between the Pearson coefficients for the BA model decreases
as the average degrees of those networks increase, as a consequence of
the small size of the adopted networks (i.e. N=100).

Bayes decision theory is the optimal statistical method for supervised
classification of data, provided the density distribution of the
characteristics of the data classes is known (It can be formally
verified~\cite{book:duda} that the Bayesian decision criterion adopted
in this work is optimal from the perspective of minimizing the
probability of misclassifications).  In the specific case of
equiprobable classes, the Bayes decision involves selecting the class
that is most probable for a set of measured properties of an element.
Suppose we have two equiprobable classes of elements, $A$ and $B$, and
let $e$ be an unknown element whose class must be determined by using
some measured property $h_e$. Provided the density distribution
functions, $\rho_A(h)$ and $\rho_B(h)$ are available, Bayes decision
theory selects the most probable class for element $e$, as that
yielding the highest value of density distribution functions at
$h=h_e$.  In other words, if $\rho_A(h_e) \geq \rho_B(h_e)$, element
$e$ is classified as A, otherwise it is classified as B.

\begin{figure*}
\centering
\begin{tabular}{cc}
\includegraphics[scale=0.20]{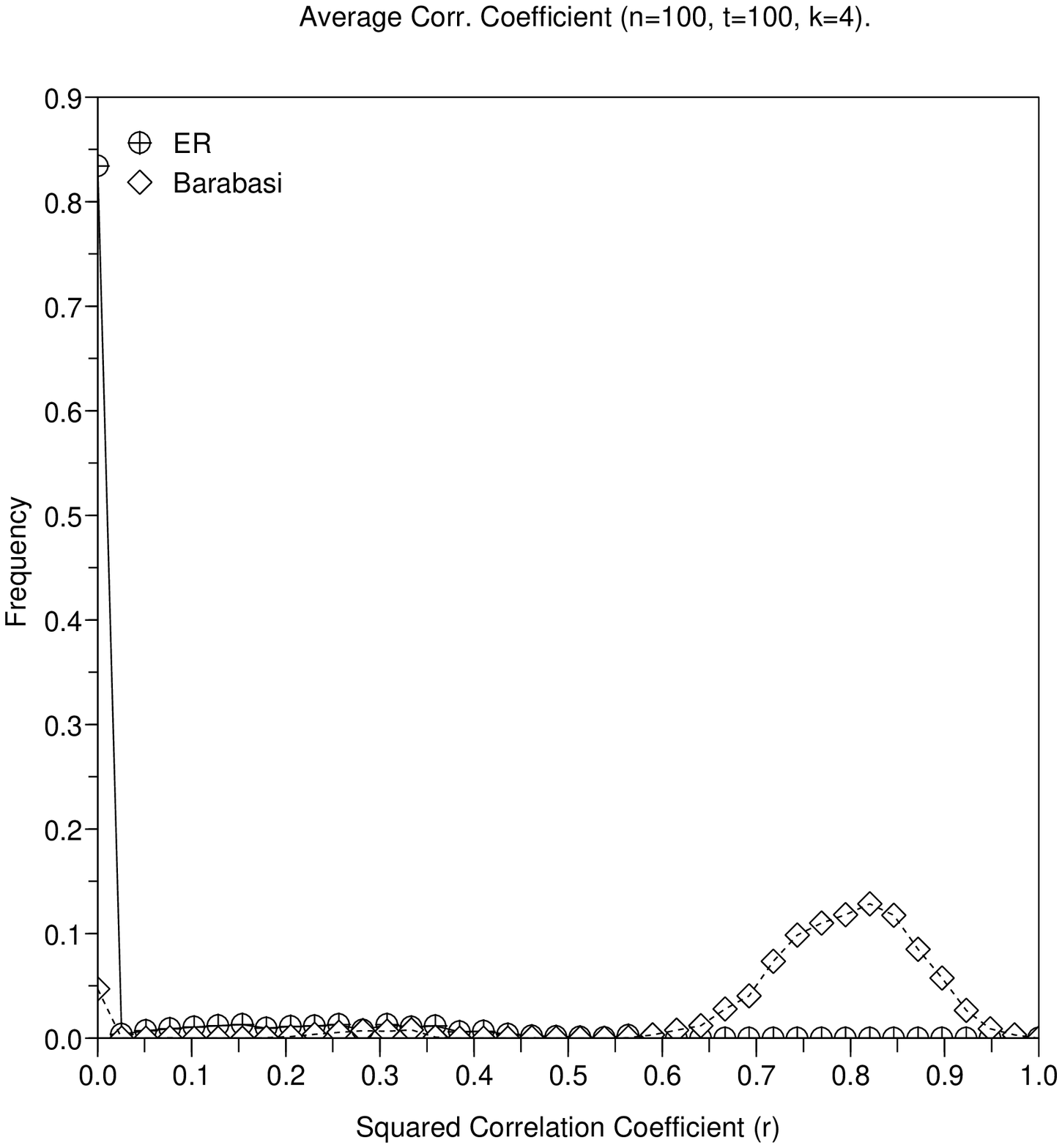} & 
\includegraphics[scale=0.20]{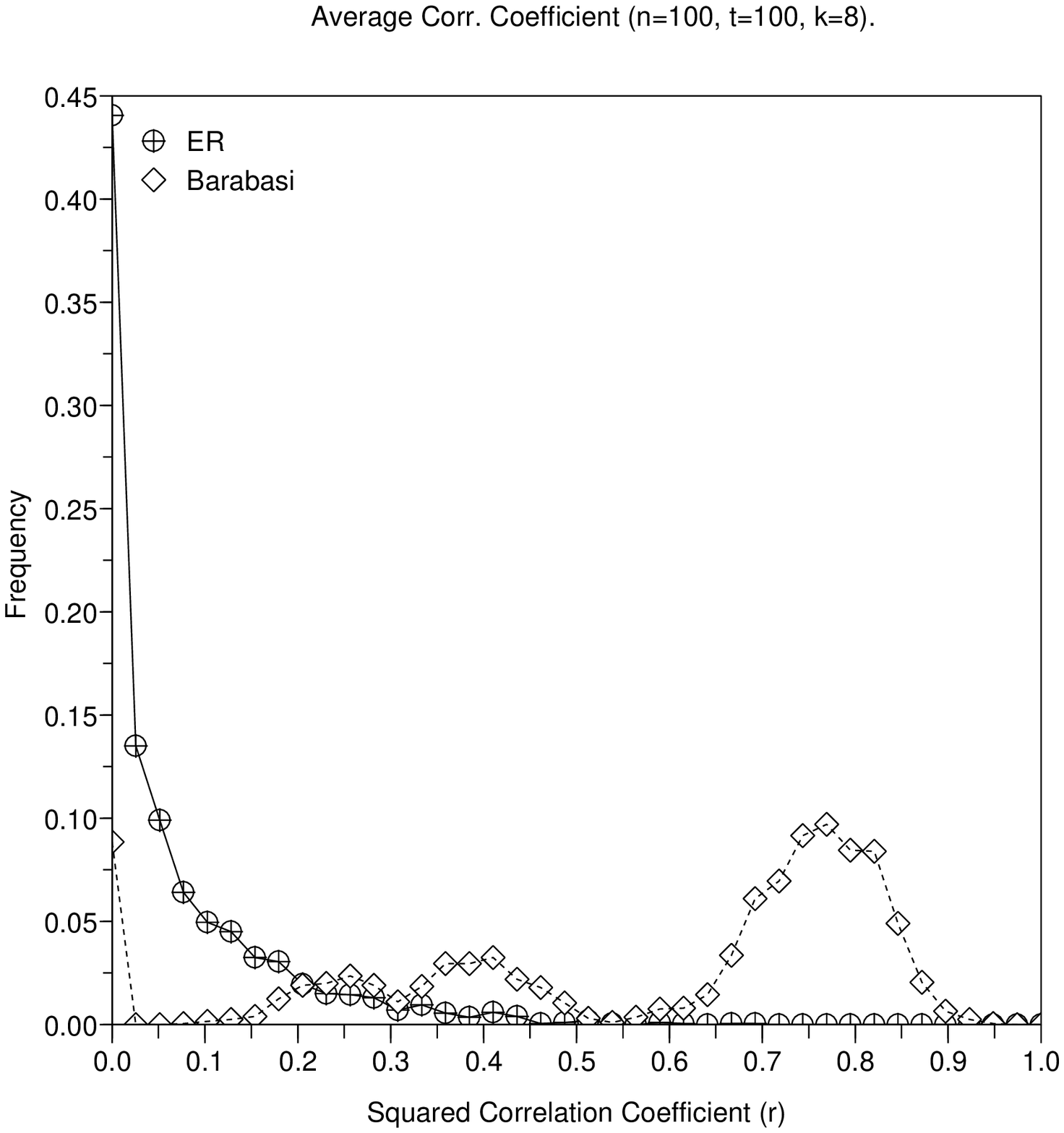} \\
(a) & (b)\\
\end{tabular}
\begin{tabular}{cc}
\includegraphics[scale=0.20]{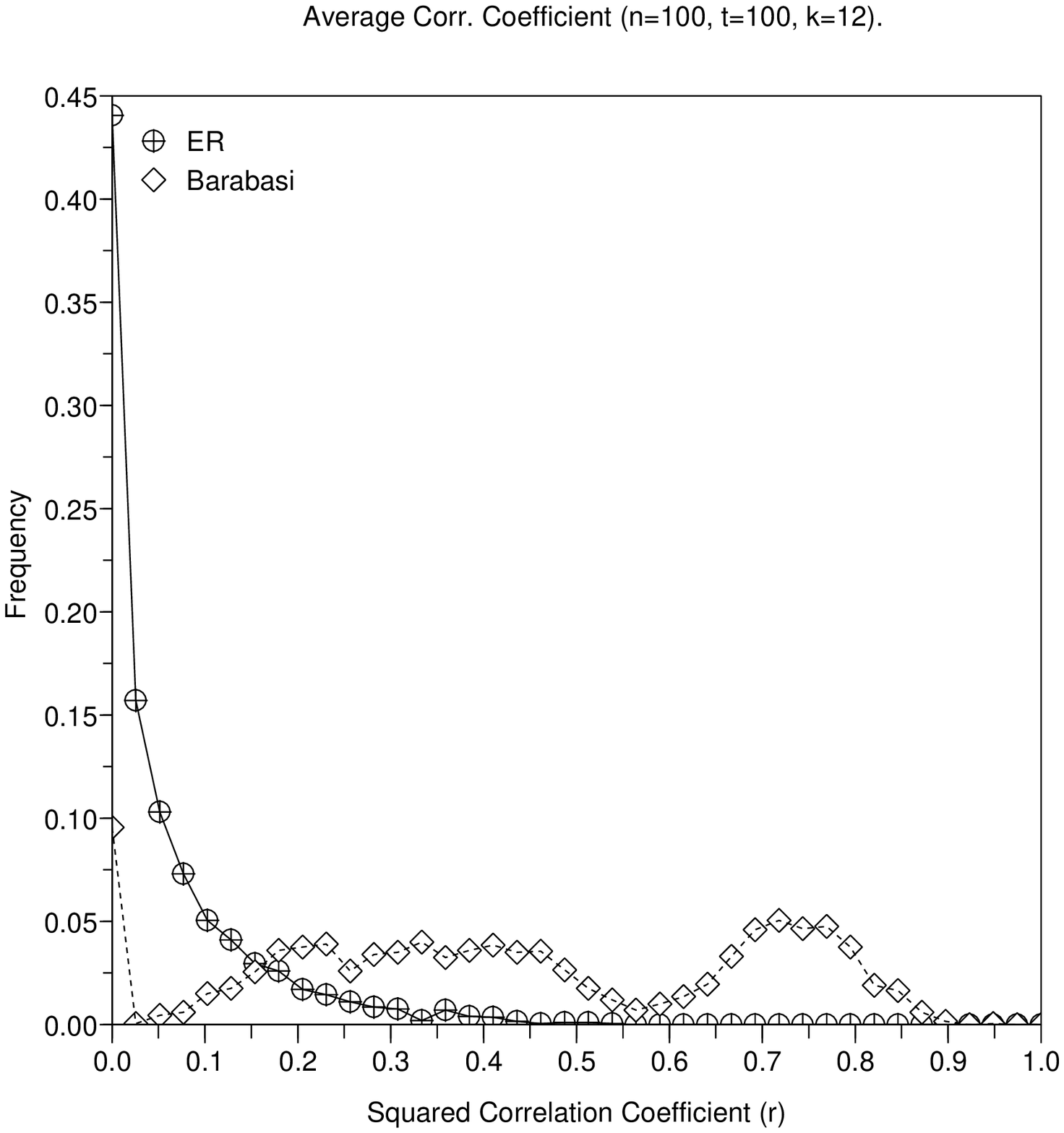} & 
\includegraphics[scale=0.20]{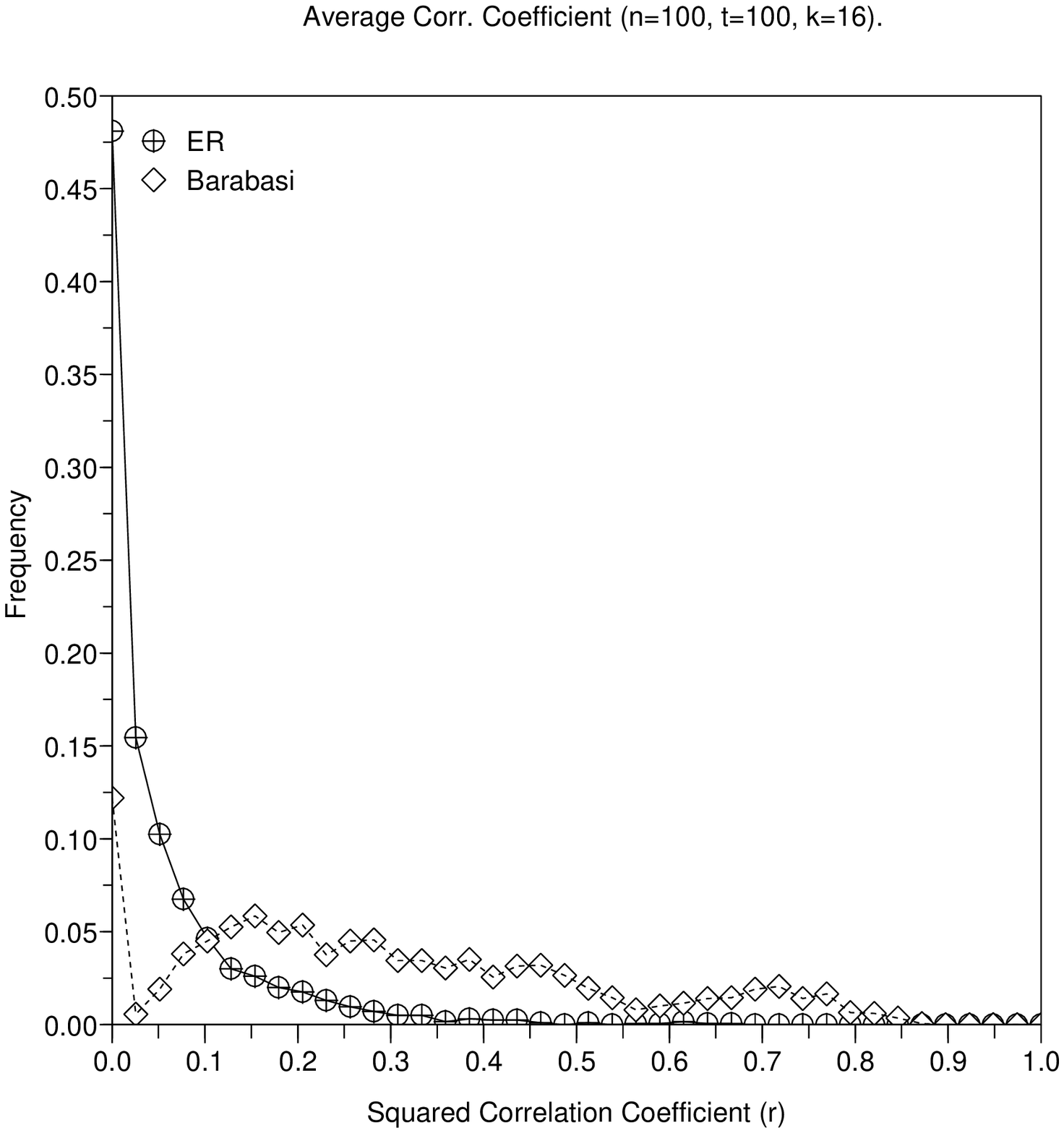} \\
(c) & (d)\\
\end{tabular}
 \caption{Average correlation coefficients for BA and ER models,
 considering every node in 2000 networks with N=100 and several values
 of $\left< k\right>$ ~\label{fig:hist}}
\end{figure*}

The Density distribution functions are not always available, but can be
estimated by using various methods.  Here we consider non-parametric
estimation from the respective normalized histogram obtained from the
measurements. In order to improve the density distribution estimation,
one can interpolate the histogram by using the Parzen windows
method~\cite{book:duda}, which consists of convolving the histogram
with a Gaussian distribution, resulting in an interpolated and
smoother curve.  Classification trough Bayes map can then be obtained
by considering the Pearson correlation coefficients obtained for the
two studied models. An example of such histograms is depicted in
figure ~\ref{fig:hist}. This figure includes the two histograms
obtained for the BA and ER models for $\left< k\right> =4$ (a), $8$
(b), 12 (c) and 16 (d). Note that the separation between the
histograms decreases substantially with the increase of $\left<
k\right>$

\section{Experimental Methodology}

\begin{figure}
 \begin{center}
  \includegraphics[scale=1.5]{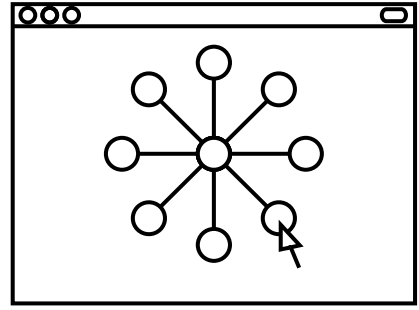}
  \caption{Navigation Screen: the subject navigates trough the network
  while trying to determine whether it is random or scale
  free.~\label{fig:tela2}}
 \end{center}
\end{figure}

A software was developed in the Java language providing compatibility
with any major operating system, specially in order to provide a
graphical interface through which the subject can navigate along
complex networks. A sequence of sets of networks with increasing
average node degree were considered.  Some parameters must be fixed
prior to each navigation, including the model of the network (ER or
BA), the number of nodes in each network, the average node degree of
the initial set of networks\footnote{A network set consists in a group
of fixed average degree networks.}, the average node degree for the
last set and the number of networks per set. The navigation starts at
a randomly selected node, which becomes the current \emph{central
node}.  The software records all the navigation actions taken by the
subject for posterior analysis.  Figure~\ref{fig:tela2} shows a
current node (center) and its immediate neighbors, where each node is
represented by a circle and the edges are represented by lines linking
these circles.

At each step, the subject is prompted to choose a node amongst the
neighbors of the central node in order to continue the walk. The
chosen node becomes the current central node, and the process is
repeated. The subject can navigate until a specific number of steps
is reached, prompting the user to make a decision about the model. No
partial results of the experiment are presented to the subject during
the navigation.

The subject can choose between either Bara\'asi-Albert (BA) or Random
model (ER). After the choice is made, it is stored and a new network
is generated and showed to subject, repeating the process for every
network of the set. After all networks in a set are navigated, the
total number of correct choices is stored and the average degree is
increased by one for the next set.

\section{Virtual Agent Navigation}

\begin{figure*}
 \begin{center}
  \includegraphics[scale=1.2]{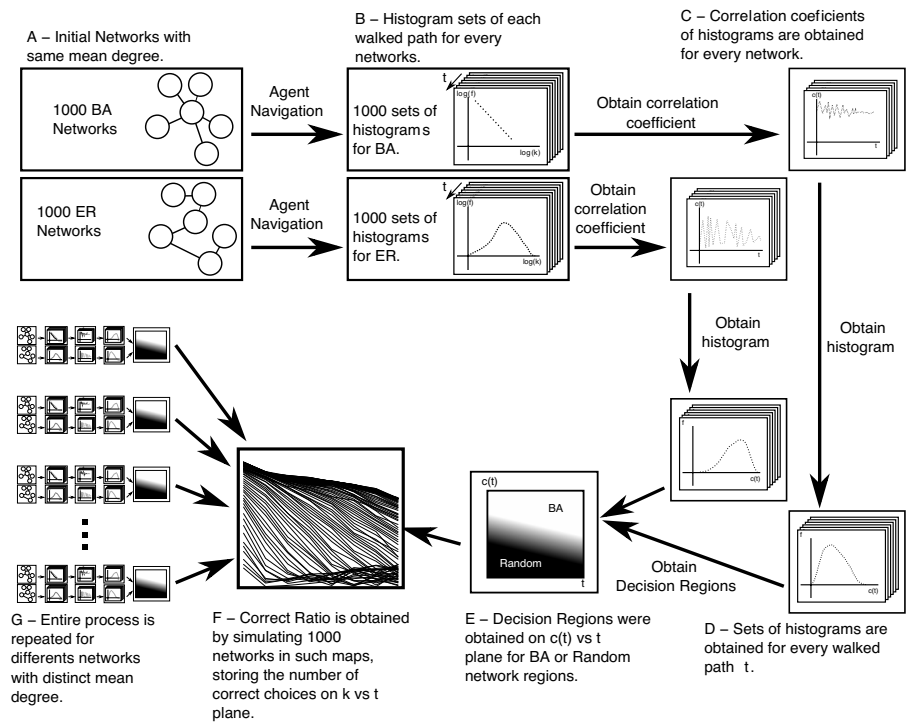}
  \caption{Schematic of main process to obtain decision regions and
  correct ratio for agent navigation.~\label{fig:circuit}}
\end{center}
\end{figure*}

In order to have a comparison standard, and also to consider
explicitly a model of navigation, an artificial agent has been
developed which is capable to navigate trough the same type of
networks as humans. The adopted heuristics is described below and is
schematized in Figure~\ref{fig:circuit}.

Two sets of 1000 networks (BA and ER) with node degree $N=100$ and the
same value of average degree are considered
(figure~\ref{fig:circuit}:A). For every network, the agent is placed
initially at a randomly chosen node and began its navigation while
using an algorithm as described in~\cite{Luciano_rand_walk}, where the
agent selects randomly a not yet visited neighbor node and makes it
the new center. If only visited nodes exists in the neighbor, the
agent use the same process to choose a new center among those already
visited nodes.  The process is repeated until every network node has
been visited.

For every new node, the number of already visited nodes, $t$, is
stored and increased. A histogram is made (figure~\ref{fig:circuit}:B)
considering the logarithmic of degree distribution along the network
at each value of $t$, from which the respective Pearson correlation
coefficient $c(t)$ is then estimated.  The same process is repeated
for every network of the same model type, obtaining a new set of
histograms $H(t)$ from the distribution of correlation coefficient
(figure~\ref{fig:circuit}:C). The same method is applied for the other
model.

\begin{figure*}
\centering
\begin{tabular}{cc}
\includegraphics[scale=0.20]{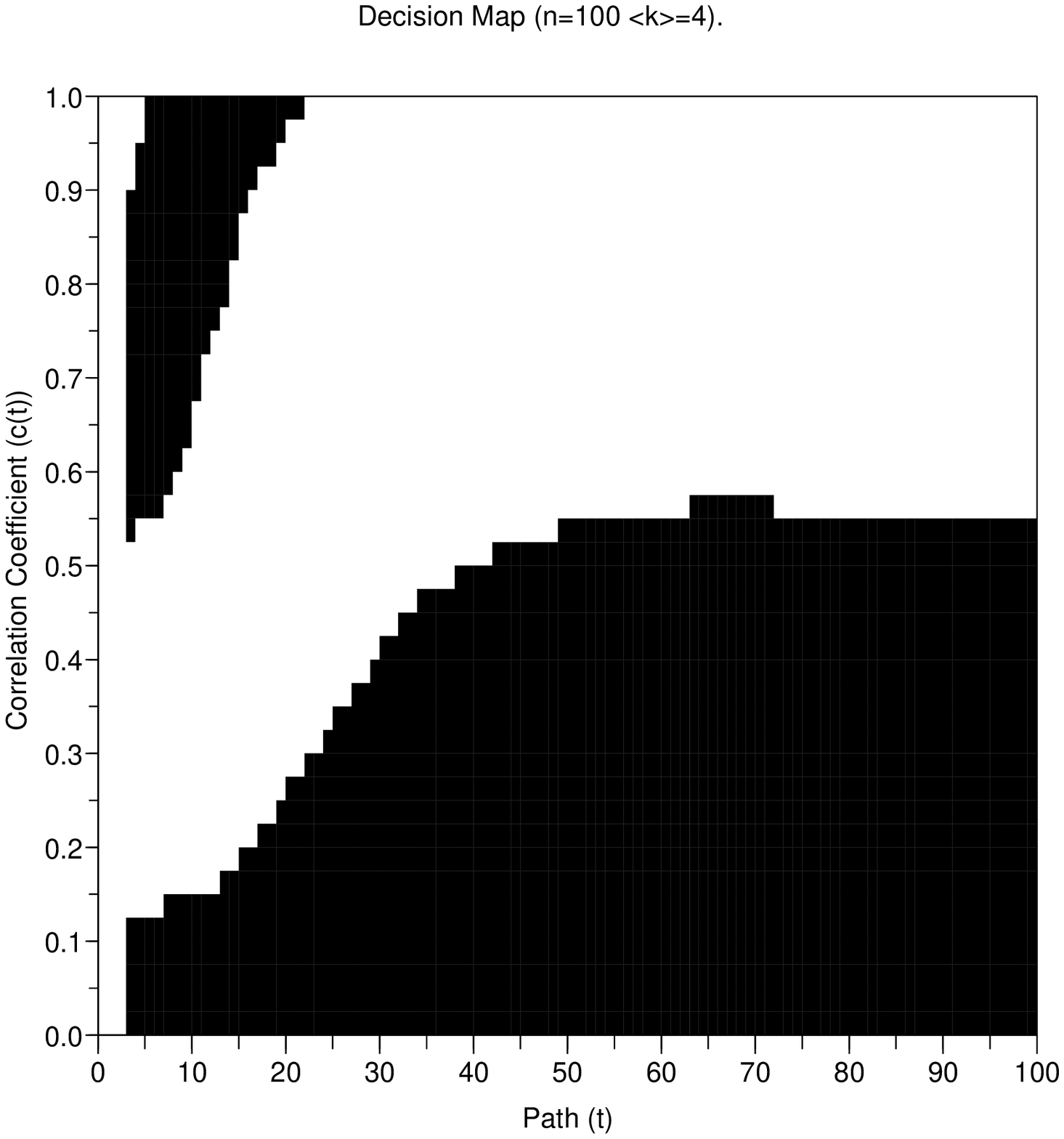} & 
\includegraphics[scale=0.20]{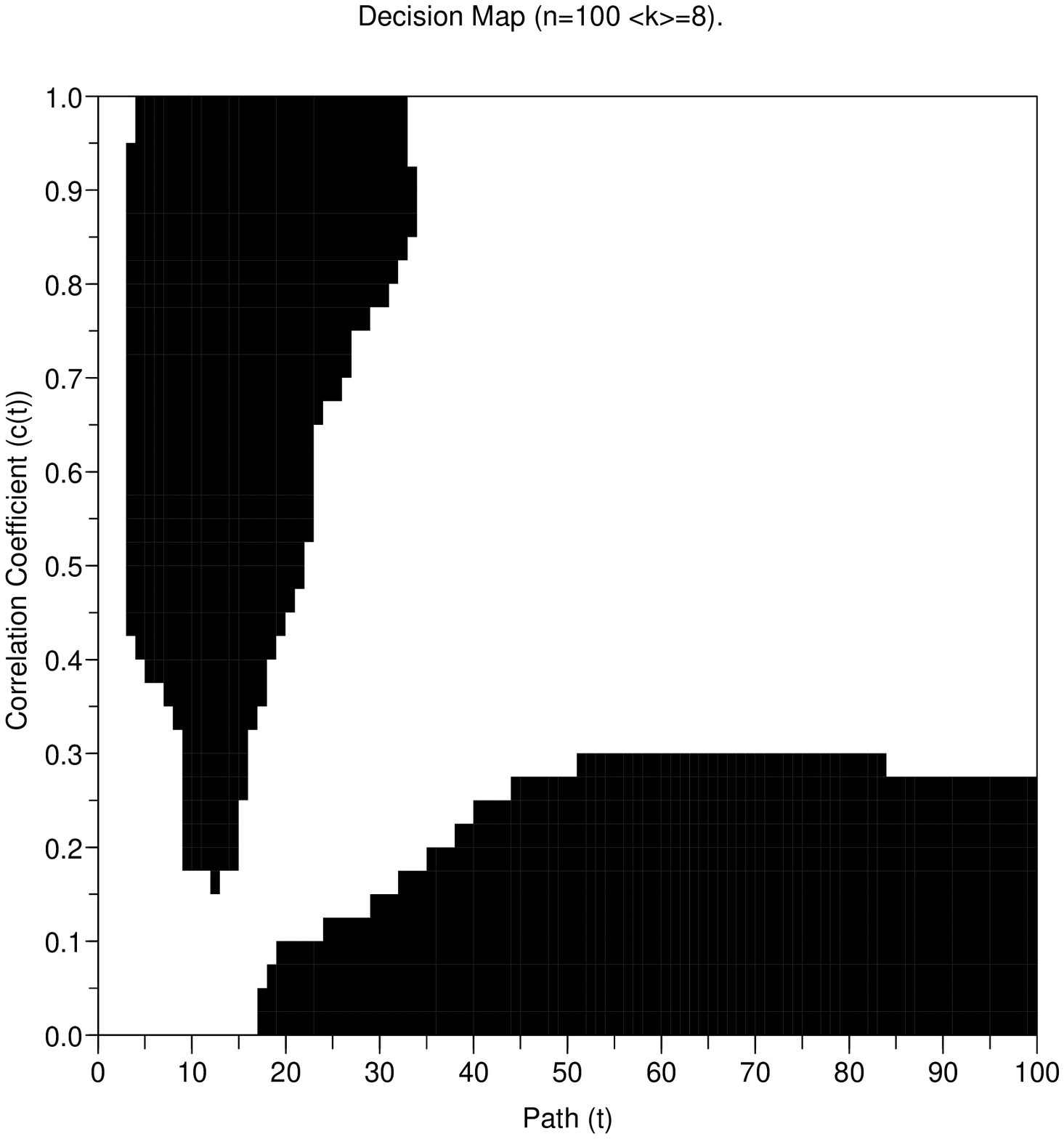} \\
(a) & (b)\\
\end{tabular}
\begin{tabular}{cc}
\includegraphics[scale=0.20]{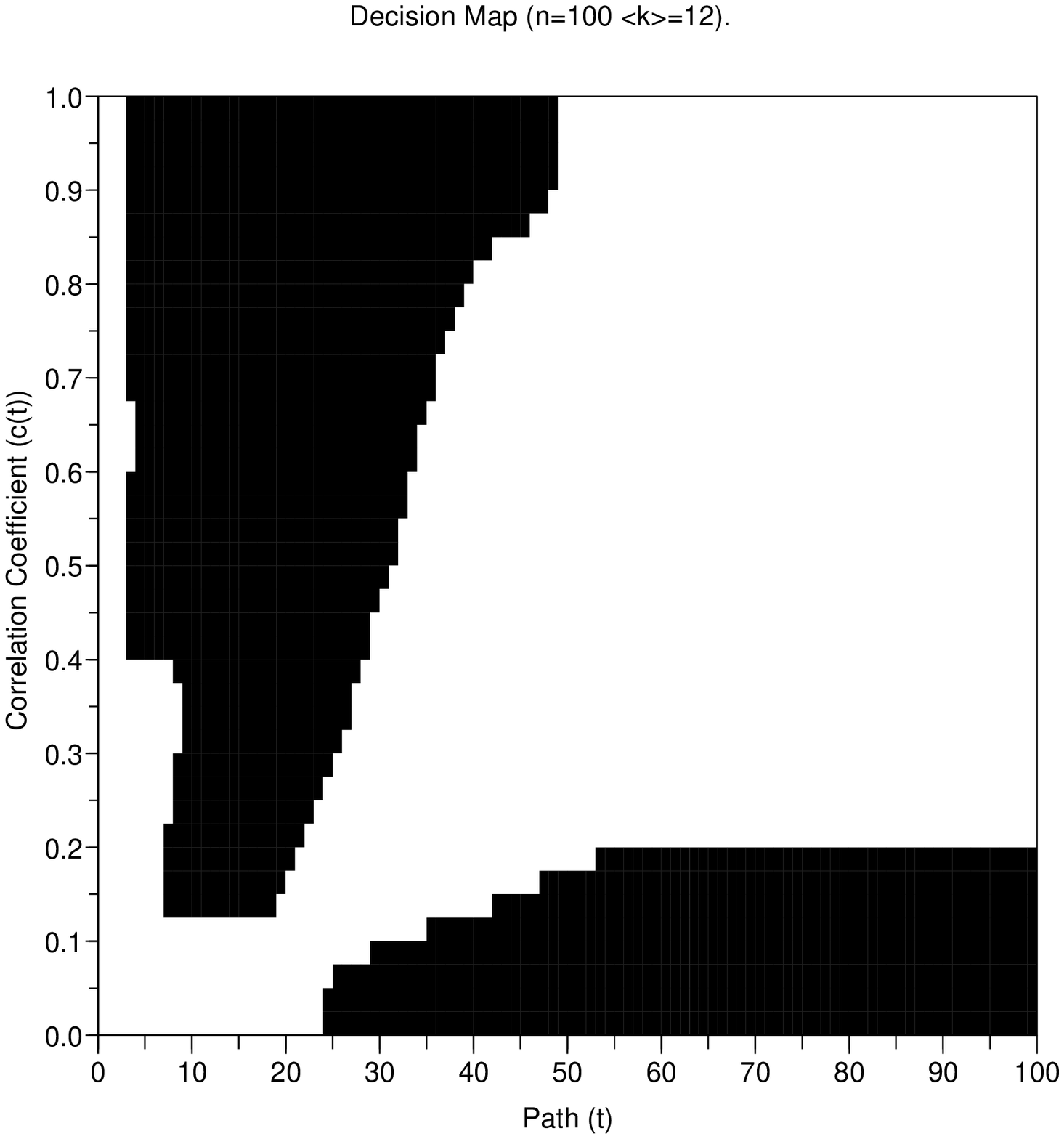} & 
\includegraphics[scale=0.20]{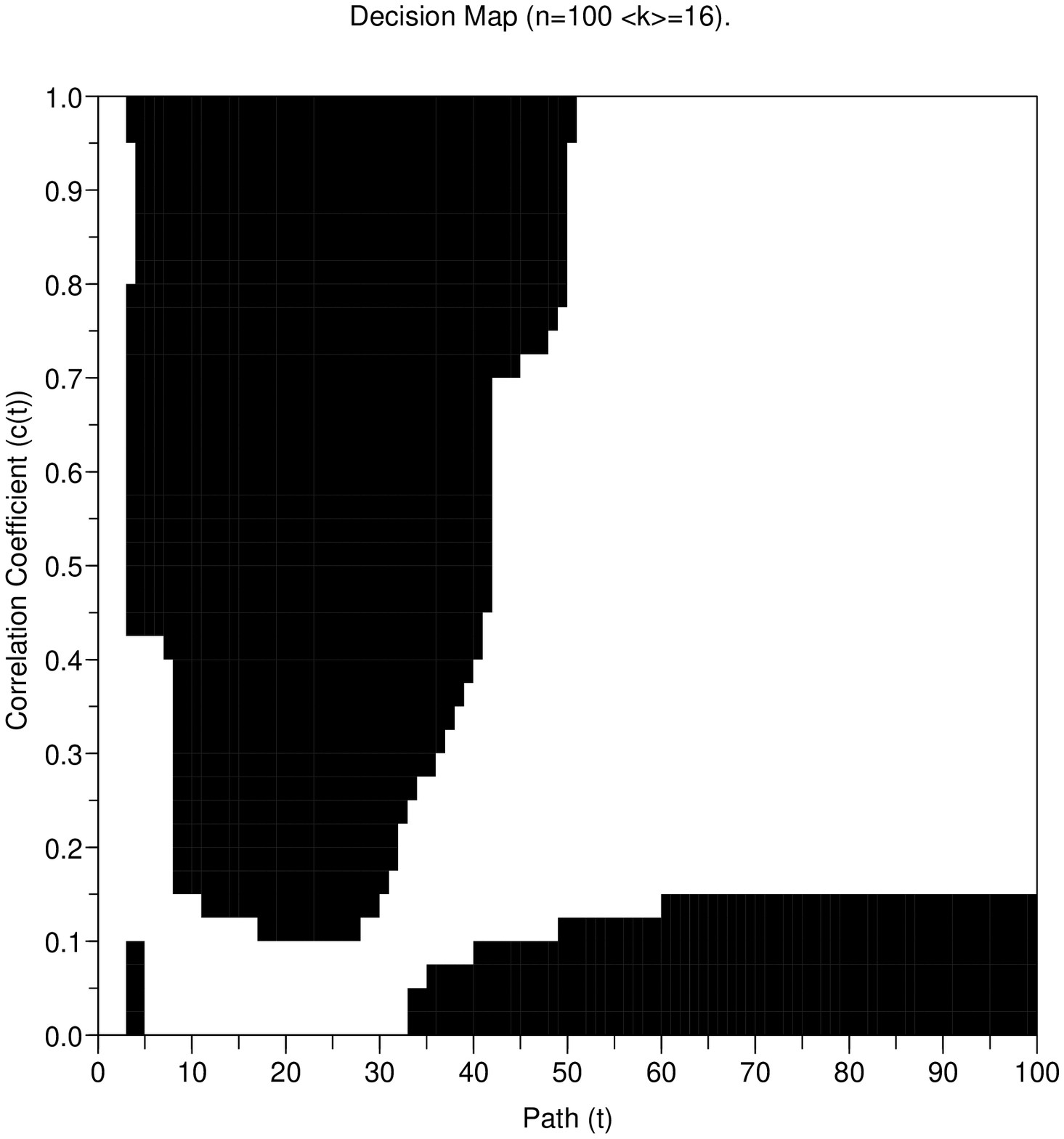} \\
(c) & (d)\\
\end{tabular}
 \caption{Decision regions obtained for the simulated models. 
  The white region represents where BA model is most probable,
  while the black region does the same with respect to the
  ER model.~\label{fig:maps}}
\end{figure*}

Because the correlation coefficients of BA networks tend to be higher
than for ER, a meaningful decision regions can be created by
considering the histograms of these measurements, indicating areas in
a plane $t$ versus $c(t)$ where the networks are more likely to
correspond to BA or ER models
(figure~\ref{fig:circuit}:D). Interestingly, different decision
regions are obtained for sets of network with different values of
average degree. A new set of networks is then navigated by the agent
while considering the decision regions, obtaining the ratio of correct
guesses for every new node visited (figure~\ref{fig:circuit}:F). Some
of these maps, obtained by performing simulations on 2000 networks of
each model are shown in figure ~\ref{fig:maps}.

\begin{figure}
 \begin{center}
  \includegraphics[scale=0.3]{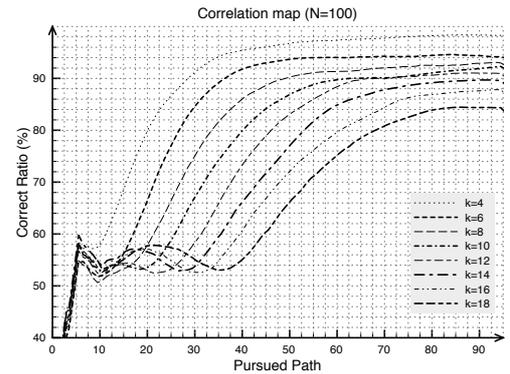}
  \caption{Correct ratio curves for the automated agents for 2000 simulated
  networks.~\label{fig:correct}}
 \end{center}
\end{figure}

The entire procedure is executed for networks with different values of
average node degrees, resulting in a correct ratio surface on $k$ vs.
$t$ plane.(figure~\ref{fig:circuit}:G). A simulation for 2000 networks
resulted in correct ratio curves by the automated agents as shown in
figure ~\ref{fig:correct} for several values of average degrees.  As
could be expected, the correct classification tends to improve along
each curve for longer pursued paths.  Also, It is clear from this
figure that the correct classifications tend to progressively diminish
for higher average degrees.  These results shows that it becomes much
more difficult to infer the nature of the network (i.e. ER or BA) when
they are more dense (i.e. higher average node degree).

\begin{figure}
 \begin{center}
  \includegraphics[scale=0.3]{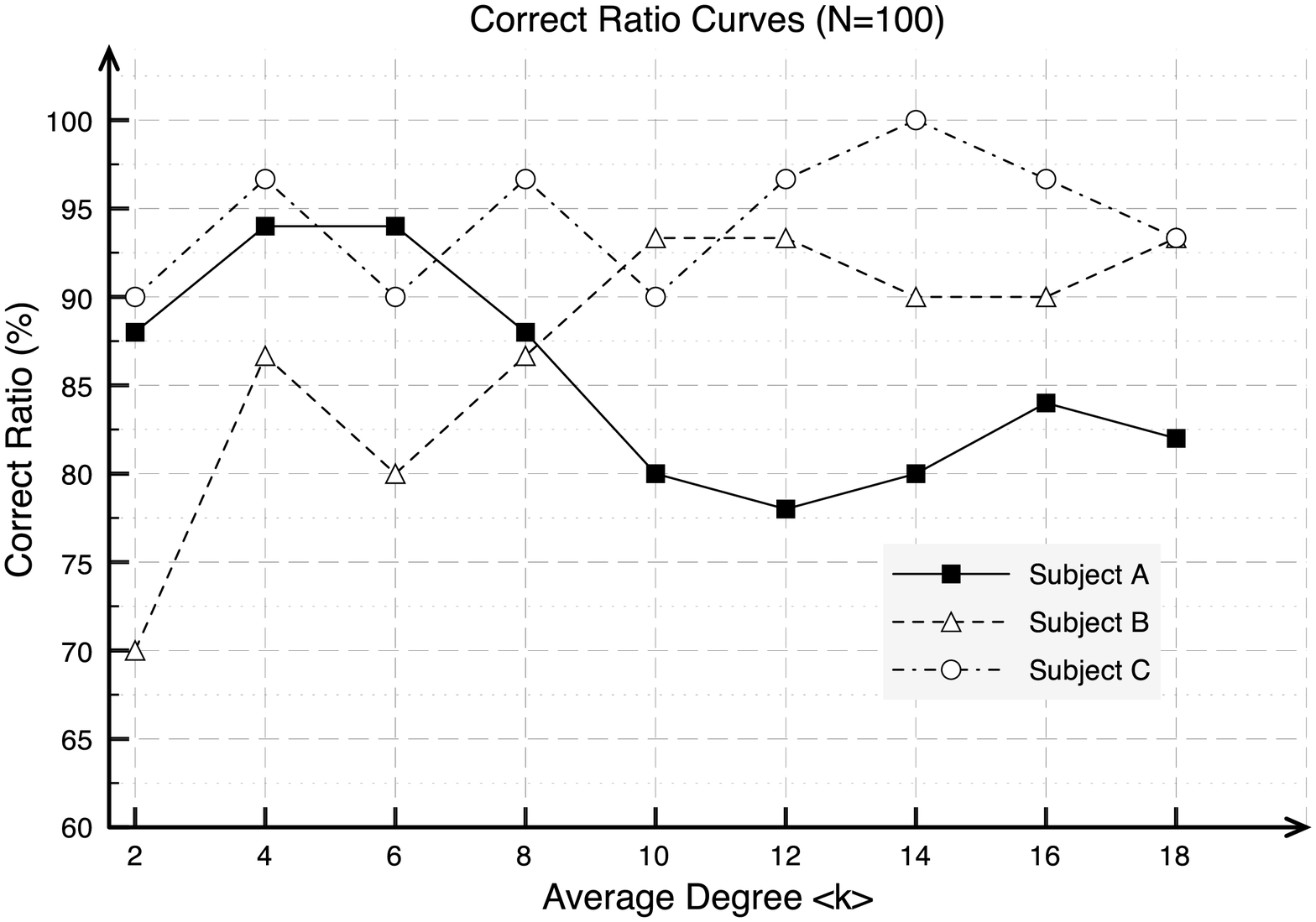}
  \caption{Several experiments obtained for three distinct subjects 
  while considering a fixed number of walks of size $15$.~\label{fig:results1}}
 \end{center}
\end{figure}

The results obtained for three subjects are shown in
Figure~\ref{fig:results1}.  This figure shows the rate of correct
classifications in terms of the average node degree.  Recall that the
classification was reached after a 30 random walks with 15 steps.
Interestingly, unlike the automated case, the performance of the
classification does not clearly diminish with the average degree.  On
the other hand, some fluctuations are observed for the correct
classification ratio.  The values of this ratio varied between 80 and
100\%.

\begin{figure}
 \begin{center}
  \includegraphics[scale=0.3]{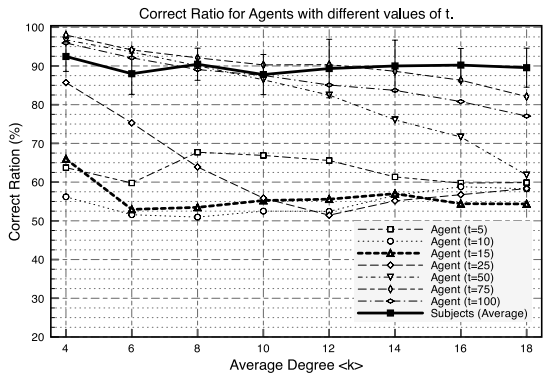}
  \caption{Autonomous agent correct ratio curves for several pursued
  paths (t) and average value and respective standard deviation for
  the results of experiments. (Comparable results are
  highlighted)~\label{fig:results2}}
 \end{center}
\end{figure}

Figure~\ref{fig:results2} shows the average $\pm$ standard deviations of the
correct classification ratios in terms of the average node degree
obtained for the automated and human agents.  In the former case,
seven curves are shown respective to different number of steps taken
by the automated agents before making the decision.  Note that a big
change takes place for the automated agents when more than 40 steps
are allowed before decision.  This changes proceeds from less than
70\% to over 85\%.  Figure~\ref{fig:results2} also shows the average $\pm$
standard deviation obtained for the humans, whose average value
fluctuates around 90\%.

\section{Concluding Remarks}

This article has investigated the important problem of deciding on the
type of network as one (automated agent or human subject) navigates
along it.  Two types of networks were considered: Barab\'asi-Albert
and Erd\~os-R\'enyi.  The measurement of the network considered for
the automated decision is the Pearson correlation coefficient
extracted from the loglog distribution of node degrees.  Bayesian
decision theory was used in order to decide on the most likely type of
network. A graphic-interactive interface was developed especially for
human navigation, with the estimation of the network type being
requested after 15 steps of the walk.  The obtained results present a
series of interesting features.  First, both the automated agent and
humans presented surprisingly good performance for identification of
the type of network.  Interestingly, such a performance tended to
reduce in the case of the automated agent, while remaining constant
(with some fluctuations) in the case of the humans.  An abrupt change
in the number of correct classification ratio was observed for the
automated agents while moving from 40 to 70 steps.

All in all, the obtained results corroborate the ability of automated
and human agents for discriminating between ER and BA complex networks
with the same number of nodes and average degree.  Additional
investigations can be performed in order to identify which topological
clues are being considered by the humans while trying to identify the
type of the networks.  In addition, it would be interesting to verify
how the consideration of additional measurements of the networks
(e.g. clustering coefficient, node correlations, shortest paths, etc.)
may contribute for enhancing the performance of the autonomous agent.

\vspace{0.3cm}
{\bf Acknowledgment:} Luciano da F. Costa thanks CNPq for partial sponsorship.


\begin{thebibliography}{0}

\bibitem{Luciano_know_walk}
  \Name{L. da F. Costa}
  \REVIEW{physics/0601118}{}{2006}

\bibitem{Luciano_rand_walk}
  \Name{L. da F. Costa}
  \REVIEW{physics/0604193}{}{2006}

\bibitem{rand_walk1}
  \Name{B. Tadic}
  \REVIEW{Eur. Phys. J. B}{23}{221}{2001}

\bibitem{rand_walk2}
  \Name{B. Tadic}
  \REVIEW{cond-mat/0310014}{}{2003}

\bibitem{rand_walk3}
  \Name{E. M. Bollt and D. ben Avraham}
  \REVIEW{cond-mat/0409465}{}{2004}

\bibitem{rand_walk4}
  \Name{J. D. Noh and H. Rieger}
  \REVIEW{cond-mat/0310344}{}{2004}

\bibitem{newman_survey}
  \Name{M.E.J. Newman}
  \REVIEW{SIAM Review}{45}{167–256}{2003}

\bibitem{barabasi_survey}
  \Name{Albert and A.-L. Barab\'asi}
  \REVIEW{Rev. Mod. Phys.}{45}{47-97}{2002}

\bibitem{luciano_survey}
  \Name{L. da F. Costa, F. A. Rodrigues, G. Travieso e P. Vilas Boa}
  \title{Characterization of complex networks: A survey of measurements}
  \REVIEW{cond-mat/0505185}{}{2005}

\bibitem{book:correlation}
  \Name{Cohen J}
  \Book{Statistical Power Analysis for the Behavioral Sciences}
  \Publ{Lawrence Erlbaum Associates}
  \Year{1988}

\bibitem{book:luciano}
  \Name{ L. da F. Costa, R. M. Cesar Jr. }
  \Book{Pattern Classification and Scene Analysis}
  \Publ{CRC; 1 edition}
  \Year{2000}

\bibitem{book:duda}
  \Name{Duda, R. O. and Hart, P. E. }
  \Book{Shape Analysis and Classification: Theory and Practice }
  \Publ{Wiley}
  \Year{1973}



\end{thebibliography}
\end{document}